# Botnet economics and devising defence schemes from attackers' own reward processes

L-F Pau, *Fellow, IEEE*

*Abstract*— This paper focuses on botnet economics and design of defensive strategies. It takes the view that by combining scarce information on the attackers' business models, with rational economic analysis of these business processes, one can identify design rules for economic defense mechanisms which the target can implement, often in a cheap way in addition to technical means. A short survey of game theory in the security area, is followed by a real case of an Internet casino. This leads to develop a model, applicable to this case and others, which is presented first qualitatively then quantitatively. This allows carrying out different analyses based on different equilibrium or termination principles; the ones studied are reward break-even analysis, and Max-Min analysis from game theory, for the target and the attackers. On that basis, a number of specific economic and information led defense strategies are identified which can be further studied using the model and specific adaptations to other data or cases.

## PREAMBLE

Needless to say, this paper does not deal with technical measures. It is not claimed either that only the suggested approaches will help. It takes a more comprehensive view rooted in exploiting some of attacker's value judgments.

## I. INTRODUCTION AND MOTIVATION

A botnet is a network that consists of compromised computers, or bots controlled by botnet masters, infected by malicious software which allows cybercriminals to control infected machines remotely without the users' knowledge [1]. Botnet owners' sources of income include DDoS attacks, theft of confidential information and application serial numbers, spam (SMTP mail relays for spam (Spambot)), IDs, financial information such as credit card numbers and bank accounts, phishing, search engine spam, click fraud and distribution of malware and adware, etc... New phishing sites are now mass-produced, with botnets used to protect sites from closure. The attackers are generally thought to be mostly motivated by self-fulfillment, fun, and proof of skills, or by specific orders to damage. However, there is every reason to believe that a significant proportion of them are also financially driven by gain, and thus develop business models that include building, exploiting botnets and trading information on targets and penetration mechanisms. A study has shown how spammers cash in, and are profitable even if they get as little as one response for every many spam mails sent out [2-4]. According to [5], botnet owners make significant amounts of money by cheating on online advertising and game agencies that use the PPC (Pay-Per-Click) scheme pay; it is claimed that about 17% of all advertising link clicks in 2008 were fake, of which a third was generated by botnets.

In addition to technical approaches, which often tend to be costly, this paper takes the view that by combining scarce information on the attacker's business models, with rational economic analysis of these business processes, one can identify some qualitative guidance on additional economic defense mechanisms which the target can implement, often in a cheaper way. While this constitutes a challenge it also highlights the need sometimes for commercial targets, to mitigate their own business models in view of those of attackers, in an equally rational way [6].

This notion of trade-off also leads the present paper to exploit the potential of game theory, although this has been too little researched in the context of botnet economics. Some papers have considered using cost-benefit analysis from the defender's viewpoint, but not exploiting the attacker's strategies other than by some statistical properties [7]. Common to the many theories in the game field, are the existence of several players having different goal functions, and of a choice of trade-offs (called equilibria) or of termination rules. While some games using quantified models for the behaviors and goals have numerical solutions, a large class can be analyzed in the presence of even uncertain parameters to devise qualitative strategies. The model developed in this study can help understanding the interaction between botnet attackers, target, and also the customers of the target.

While acknowledging that insights into attacker's strategies are few, this paper exploits the observation and partial information about a case in the Internet game area, falling initially under the category of gain from click fraud.

The interesting aspect is that this case represents a "real honeypot" as opposed to "virtual bots" used sometimes. This case also shows that attackers can to some extent organize themselves as economic agents with different roles, such as botnet masters and attackers who rent bots from the previous. The gain-driven botnet masters and attackers make decisions on parameters of the efforts and penetrations, such as the optimal size of botnets, botnet operating costs, the bot rental prices and assumed relative skills between them.

The paper is organized as follows. After a short survey of the limited research on the use of game theory in the security area (which applied readers may jump over), is presented the case of an Internet casino. This illustrative example brings us to develop a model, applicable to this case but of much more general nature, which is presented first qualitatively then quantitatively, with later simplifications. This allows to carry out, as explained above, different analyses based on different equilibrium or termination principles; the ones studied are reward break-even analysis, and Max-Min analysis from game theory, for the target and the attackers; additional simplifications allow to make more explicit some specific results including using data from the case. Before proceeding with summarizing suggested defense mechanisms, criticism of the model is made on the basis of day-to-day observations. A number of specific economic defense strategies are identified which can be further studied using the model and specific adaptations to other data or cases.

## II. OVERVIEW OF SECURITY GAMES

A game is a contest wherein several players choose, according to a specified set of rules from among a number of permitted alternative actions, in an effort to win certain rewards. In the security area, the players each have assets to protect, and rewards to win amongst themselves and others. Such games are usually non-cooperative, and furthermore dynamic, and sometimes involve the formation of coalitions; there are very many theories and variants, and the reader is referred to [8,9] for comprehensive relevant exposés. Players make their decisions independently and each one seeks unilaterally the maximum possible gain, of course by also taking into account the possible rational choices of the other players. A special case, called Max-Min (or Min-Max) in gametheory, is when each player maximizes unilaterally his gain when the other players minimize simultaneously their losses (or maximize their own specific gains). Another equilibrium type is Nash, where no player unilaterally can win by moving away from the equilibrium if it exists.

In [10], it is assumed that attacks are on one target at the time; a certain amount of effort is put progressively, with increasing and convex costs, but attacker only achieves a probability of success in yielding a reward. The attacker will seek to maximize the actual reward, but has to decide as a strategy on an optimal stopping rule by computing the expected gain of carrying on with the attack and carrying the costs. The model of Section 4 is a variant but takes the pragmatic view that the probability of success is only skills and time based, and not cost related. Reference [10] gives some results assuming an explicit form $(1-\exp(-t/\alpha))$ for the probability of success with time t, and an explicit linearly increasing cost with time. It also considers the case of switching costs in changing target, if the expected benefit of going on with the current target gets below the expected benefit from the new target. Assuming the attacker realizes the target's security level after switching to it, he will both pay the switching cost and higher penetration cost; this implies that a target should apply in cycles increasing security levels to deter attackers. The target may also claim to be highly secure, and thus the attacker does not know whether the target has a high or low security level; the attacker is then forced to apply Bayesian inference, and it is shown that low security targets are better off hiding the fact that they are low-security, and high security targets should advertise that they are highly secure.

In [11] was studied a non-detection "cat-and-mouse" case as a Nash equilibrium "game, with malicious packets hiding within a normal flow, sent by an attacker trying not to be detected, while the defender is active. The attacker must select a path, such as a highly loaded link; to send his malicious packet so as to minimize the detection probability .The defender must select the links to scan to maximize the detection. In [12] the reverse is studied in a similar way, in that the defender wants to send some flow through a network with vulnerable links subject to attacks.

## III. CASE: ATTACKS ON AN INTERNET CASINO

### A. Context

The observations below are from a specific context, and obviously cannot be generalized. The target is interesting and viable for attackers. But it also represents a site needing monitoring and control by public authorities under evolving legislation for Internet casinos. The host country for the Internet casino was one of the first in Europe to allow on commercial and regulated terms such games for real money. Furthermore, because of the permeability between players in real casinos and those trying out Internet casinos, as driven by expertise in the specific games offered, some information could be collected about the comparative expected rewards, and thus on the reward levels expected by attackers. Finally, traffic monitoring could be effected with the assistance of the ISP meeting the conditions of the evolving legislation and procedures applicable to Internet casinos.

### B. SIP vulnerability

A problem with single sign-on with SIP opens up for malware, and allows for botnet attacks at the SIP signaling level. This despite IETF protocols for intrusion detection (IDXP, IDMEF; RFC 4765-4767) which can stop botnets close to the root. It allows botnet control traffic to be tunneled inside legitimate SIP clients, and thus for attackers to carry out learning on a target.

*C. Exploitation by attackers and observed data*

Viewing botnets as a "service" exploited as indicated above, attackers rent the access to the SIP network among themselves on a time basis, in view of carrying out the learning and the attacks represented by a botnet "success". Such a "success" is defined as taking over the target's application, here the casino, with low observability, and minimal command-and-control traffic, while collecting revenue from renting the botnet for spamming and/or or intrusion, until detected. In other words, the attackers want to take over and use the casino in a stealthy way. The revenue from the intrusion depends on the attacker's ability, while in control, to siphon off assets under control of the target. Normally, the Internet casino receives the bets from its customers via a third party deposit bank, which is out of reach from attacks. However, the Internet casino itself decides on payouts to winning customers. Assets over which control may happen are the orders to execute payment of game wins to winning customers, and possibly to declare as winners all players on a given game at the time of the intrusion, with payments to be effected via payment orders to the bank.

The typical rent until "success" was reported to be in the 50 000 USD range for maximum 24 hours, after which the party renting out access to the site in the way described would terminate the sign-on problem, and thus deny the tenant attacker to continue in the quality of botnet master. Due to constraints on game session durations and max wins per game, the absolute ceiling for cumulated rewards to an attacker from intrusion over 24 h has been estimated by the casino company to be about C= 2800 USD.

This leads to the conjecture that attackers sometimes earn income from each other, supplementing opportunistic rewards from successful attacks on the targets. They may use a kind of arbitrage. It has not been investigated how attackers communicate with each other; nor is it known if the rental market is organized or not, besides bilateral deals which the model below is assuming.

A technical research issue is to build testbeds to analyze botnets, open stack real-time network emulators such as CORE [13], to determine what is the lowest bound on command-and-control traffic needed to maintain a botnet running, and thus use this information to deny botnet attacks for given threat models.

## IV. MODEL

*A. General case*

The situation depicted in the case, as well as many others, can be formalized as a game between (N+1) players, where player i=0 is the target, and players i=1,…,N are the attackers. The reward structure for each player "Reward (i)" is a money flow, meaning the total cumulated income (or cost) over a time period T.

The Reward (0) to the target (0) is:
- gaming income: proportional to the product of total number of third-party paying customers ($\int n(t)\,dt$, t: 0 to T), by the fee per game session "usagefee";
- instantaneous cost of winning customers: proportional to the average winning probability pwin to the instantaneous number of customers n (t), by the ceiling "C" on rewards to the winning customers per game session.
- intrusion cost: the cost of successful intrusions to the target (0), as specified below; the attackers are supposed not also to be legitimate customers.

The probability of a reward to an attacker (i) by a successful intrusion of the target (0) is a time-dependent probability function "p (i,t(i))" corresponding to the learning process involved in possibly achieving a reward after a lapse t(i). The reward "Reward(i)" to one attacker (i) is the sum of the income:
- from rentals from other one's (j) is a deterministic function of time "rent (i,j,t(j))" and the sum of these rental incomes from all other attackers j;
- from intrusions: assuming the attacker (i) to take control of the target (0) after a lapse t(i) after he has become a tenant, and thus to syphon off all maximum rewards otherwise due to customers, his revenue will, in the best case for him, be the ceiling on rewards "C" multiplied by the number of customers n (ti+t (i)) at that point in time.

As to legitimate game customers, the net revenue to a given paying customer Reward (customer) at time t, is his winning reward (pwin*C) minus his "usagefee" per session.

At any time, due to the nature of the vulnerability described in Section 3, only one botnet attack can take place, once a player (i) has granted a different player(j), the access against the payment of rent(i,j,t(j)). Thus, in a general case, a Markov model is needed to represent all the transitions between attackers i=1,…,N, each operating for a lapse t(i), apart from lapses with no on-going attacks.

*B. Simplification with equal attack durations*

A simplification, sufficient for the subsequent design of strategies, is to assume that all attackers carry out multiple simultaneous attacks for the same lapse t(i)=Δt, and that there are no idle periods; the number of attacks shall then be A=N(N-1) and the average duration average(Δt)=T/(N*(N-1)). The time ti at which attacker(i) starts his attack then becomes a specific multiple of Δt, dependent on the order of the attackers. We then get over time period T:

$$\text{Reward (customer)} = pwin * C - usagefee \quad (1)$$

$$\text{Reward}(0) = (usagefee - pwin) * [\int n(t)dt, t=0, T] - C * [\sum n(k*\Delta t), k=1, A] \quad (2)$$

$$\text{Reward}(i) = \sum ([rent(i,j,\Delta t) - rent(j,i,\Delta t)], j=1,...N) + p(i,\Delta t) * C * n(t_i + \Delta t) \quad (3)$$

$A = N*(N-1)$     (4)

$\Delta t = t(i) = t(j) = Average\ (t(i)) = T/[N*(N-1)]$     (5)

The best intrusion reward assumption to an attacker can be changed by modifying the ceiling C.

*C. Generalizations*

-The model above is quite adequate to describe a wider class of botnet economics where the Internet casino case is replaced by a broader class of services to customers from which the defender collects revenue, part of which may be exposed to attackers. What would change is the "Reward (customer)" part possibly even disappearing to be replaced by quality of service satisfaction goals; the income and cost parameter parts of Reward (0) would also have to be adapted.

-The model already incorporates the case where the attacks of some of "i" are for pay, in that "rent(i,j,t(j))" can be extended to the case where rent(i,i,t(i))≠ 0 representing the payment to "i " received by a paymaster. Likewise, it is easy to give special characteristics to the business parameters of a botnet master compared to other attackers.

-Also, it is easy to incorporate in Reward(i) an additional term representing attack costs to attacker "i"; they have in this paper been neglected as no data where known in the case depicted on the cost structure of the attackers.

-Finally, under suitable distribution assumptions, likelihoods of rewards and savings can be estimated.

## V. ANALYSIS

The model and data of Sections 3 and 4 allow studying behaviors under different types of assumptions and concepts.

*A. Break-even analysis*

This corresponds to the behavior of the players when they just break even over a time period T, that is their rewards are 0, as their income is outweighed by costs.

-a1. for attacker "i":

$p(i,\Delta t) = (\sum([rent(j,i,\Delta t)-rent(i,j,\Delta t)], j=1,...N))/(C*n(ti+\Delta t))$     (6)

which tells that the attacker must, to achieve breakeven, maximize his probability of success from learning, by negotiating as asymmetric rents as possible with other attackers, and hope for the target to have as few customers as possible at the time of his attack. This means that in-fighting / competition between attackers is for them a necessary condition for break-even .It also confirms from the condition $p(i,t) \leq 1$ that:

$\sum([rent(j,i,\Delta t)-rent(i,j,\Delta t)], j=1,...,N) \leq (C*n(ti+\Delta t))$     (7)

i.e. that the rent differences are bounded by a function proportional to the maximum reward and to the number of customers.

Assuming the rents rent(i,j,t(j)) to be proportional to their duration so that rent(i,j,$\Delta t$)=rent(i,j)*$\Delta t$, the same inequality, when data from Section 3 are entered for "rent" and "C= 2800", and "$\Delta t$" is replaced by its value, shows that there is a lower bound on the number of attackers N which increases roughly proportional to the duration T, and decreases when the number of instantaneous customers grow; this is a normal consequence of the learning process:

$N(N-1) \geq 17* (T/n(ti+\Delta t))$     (8)

In all cases, there is an intrinsic contradiction in benefits to the attacker aiming for breakeven, between a target with few and many customers.

-a.2: for the target i=0: Rewards(0) are:

$(usagefee-pwin)* [\int n(t)dt, t=0, T=C*[\sum n(k*\Delta t), k=1,A]$     (9)

which means that that the target must for breakeven increase the "usagefee" charged to customers with the expected number of attackers/attacks, and obviously with the maximum reward to customers. In the special case where the instantaneous number of customers is constant n(t)=n, the corresponding simplified breakeven condition shows a quadratic dependency on the number of attacks:

$usagefee = pwin + C*N*(N-1)/T$     (10)

*B. Max-Min analysis*

The well-known Max-Min equilibrium, or saddle point, in games with one defender and many attackers, aims at determining the probability distributions of the decision variables of the defender and attackers, so that, simultaneously, the defender minimizes his average loss and the attackers maximize their average gain.

In the case of the attackers, the key decision variable is the duration of each attacker's rent and learning period "t(i)". In the case of the target, the key decision variables in the presence of unknown number of attackers and unknown number of customers, is how to set the "usagefee" and the maximum payout "C".

This game is fundamentally not a zero sum game, term used to depict the circumstances where the loss to the target, is balanced off by an equal gain of the attackers, and vice-versa. Thus one must separate out the game strategies driven by either the target's concerns or the attacker's concerns.

As this paper deals with the attackers, in view of devising weaknesses in their strategies, the Max-Min strategy of an attacker i=1,..,N will be such that it satisfies:

$Max[Reward\ (i) | t(i)]\ (Min[Reward\ (0) | (usagefee, C)]$     (11)

where | stands for a condition on subsequent argument being given. As the inner-loop maximum can be shown to be

linear with a positive coefficient w.r.t. "usagefee" and negative w.r.t. "C", we must assume a maximum value called "usagefeeMax" and a minimum value for "C" called "CMin", so that the Max-Min strategy of the attacker i satisfies:

*Max [∑([rent(i,j,Δt)-rent(j,i,Δt)],j=1,..N) + p (i, Δt) * CMin\*n(ti Δt) | t (i)]* (12)
*CMin ≥ usagefeeMax ≥ 0* (13)

which shows that, such a strategy is satisfied when the attack duration t(i) maximizes the probability of a reward, meaning when the efficiency of the attack is maximum; this is an obvious result. If that Max-Min result does not achieve the break-even condition of Section 5.a1 for C=CMin we are back to that case.

The Min-Max strategy of the attacker would correspond to the guaranteed minimum reward to attacker (i) when the target maximizes his reward, that is:

*Min [∑([rent(i,j,Δt)-rent(j,i,Δt)],j=1,..N) + p (i,Δt) * CMin\*n(ti+Δt) | t (i)]* (14)

which corresponds to the case where the probability of an intrusion reward is minimum, including zero. It can be seen that even in that case, the attacker (i) may still achieve a positive minimum reward, either because the rent structure between attackers is skewed in his favour, or because the attack learning efficiency is still non-zero positive and the minimum reward CMin is high.

In the special case where the instantaneous number of customers is constant n(t)=n, the corresponding simplified Max-Min and Min-Max conditions relate to the optima over t(i) of the expression :

[∑([rent(i,j)-rent(j,i)],j=1,..N) + p(i,Δt) * CMin\*n\*N\*(N-1)/T (15)

which shows a very complex and interesting interdependency. Even with certainty on success, the attacker "i" can still achieve a negative Max-Min reward if the rent structure between attackers, their number, the number of customers and the minimum reward satisfy jointly a negative value for the expression (15).

## VI. STARTING FROM PRACTICE

It can easily be claimed that the analysis made in the preceding Sections is theoretical and remote from some day-to-day operational observations which are summarized below, but these observations in turn give no clue as to systematic strategies as suggested in this paper in Section 7.

Some assumptions are made about a general attacker's (i.e. bot herder's) course of action and that maximizing an immediate financial reward is their ultimate goal. Observing general botnet attack cases, one may conclude that attackers do not organize their resources with regards to professional economic strategies. Providing a setup of virtual bots to introduce uncertainty and added costs for the botmaster implies effort on the defenders' side too. Additionally, it is arguable if attackers would even care or just go on with some virtual bots that cannot be used for attacks.

It is suggested to introduce a virtual attacker that exploits the cooperative structure of the attackers renting services from each other, by influencing the pricing standards, thus influencing the market in a way that some services are simply not profitable anymore. This means that an open market is assumed as random people are allowed to join, which may not always be true. On the other hand, the attackers do not want a closed market, as then it would be easier to infiltrate the attackers' communication channel to track them down and prosecute them.

The model assumes that the victim of an attack has some degree of freedom for controlling the availability of his service. In the case presented, the online casino would buffer user communication and process the queue in bursts. By varying the input processing strategy, an attacker's reward strategy shall be influenced in a way that maximization is more difficult. This paper cannot provide evidence that attackers act on such a formal level. At the same time, the assumed flexibility is not given in many cases, but can be engineered with clear advantages.

## VII. ECONOMIC DEFENSE STRATEGIES

Existing technical approaches aim, either at to preventing infected machines from reaching the target, or to redirect the visit of infected computers to a different site. Such defenses tend to be passive, costly and sometimes inefficient mainly because current Internet architecture makes it extremely difficult to differentiate a "pretend-to-be-legitimate" request from a "true legitimate" visit.

That is why this paper proposes to identify some additional economic and information based defense mechanisms , by combining scarce information on the attackers' business models, with rational economic analysis of these business models, which the target can implement, often in a cheaper way. Most of these mechanisms aim at creating economic dis-incentives to the attackers. We list below those identified in this paper, mostly from the analysis in Section 5:

### A. Virtual bots

The idea of using virtual bots (honeypots running on virtual machines that can be compromised by the botnet masters) to create *uncertainty and added costs* in the level of botnet attacks is known [7, 10]. The uncertainty introduced by virtual bots has a significant impact on the profits gains from botnet attacks; it generates decreasing profitability, mostly by increasing real or estimated costs, and reduces the botnet related activities by economic dis-incentives. At any point in time, the capacity of the server limits the number of supported compromised machines, further limiting the number of bots rents used to attack the target.

### B. Maximizing the asymmetric nature of botnet rents

We have shown above that the differential term $[\sum([rent(i,j, \Delta t)-rent(j,i,\Delta t)], j=1,..N)]$, representing the rental income to attacker "i" from all other attackers, plays a major role in break-even as well as Max-Min equilibrium analysis. This means that a virtual attacker controlled by the target, can drive down the rewards to the attackers, even in case of successful penetrations, just by choosing a suitable rent structure amongst the attackers, which this virtual attacker can advertise on suitable channels.

### C. Using customer traffic variability

We have shown above that the success probability from learning "$p(i,\Delta t)$" is severely affected in the break-even and equilibrium conditions, by the instantaneous customer traffic density $n(t)$ in relation to the total customer volume over the time period T. For services which are not too time sensitive, such as those from the casino case (and many others), it pays off to periodically buffer customer traffic to create, via fluctuations in how $n(t)$ affects the rewards, such fluctuations in the learning periods $t(i)$, so that attackers "i" will consider the learning to be ineffective.

### D. Customer traffic fluctuations affecting dynamically possible botnet rents

We have shown above that the rents are bound by an upper bound proportional to the customer traffic density. If that bound is operative, as it is likely to be as attackers will all each try to maximize their statically defined rental incomes, just having to renegotiate rents dynamically creates such an overhead to the attackers that they will skip them.

### E. Determining carefully the minimum reward and payment time

We have shown above that in the Min-Max equilibrium case, the guaranteed reward to an attacker may be low, if not even negative, if the product $[CMin*n*N*(N-1)]$ is low. The number of attacks shall be kept low by technical means, but the disincentive created by minimizing the on-line real-time rewards is likely to be syphoned off by a penetration. While respecting applicable regulations, in this case to casino rewards, the defender can in his on-line systems run a very low real-time reward CMin, while instructing the bank to apply C to time-shifted payments to legitimate winning customers.

### F. Adaptive economic defense strategy

We have shown that, under simplifying assumptions, the Max-Min and Min-Max rewards to an attacker (i) depend on the top or floor values of the expression in Eq (15). In the case where the defender may have an estimate of the attack density from detection probes, he can adaptively modulate n and CMin to confuse the attacker's perception of his success probability and reward level.

### G. Claiming high security protection

Reference [10] suggests that, under some theoretical assumptions, a target is better off hiding the fact that it is low-security, and a high security target should advertise that it is highly secure. This paper does not claim that this approach will work, but only that it has been suggested and tested.

### VIII. CONCLUSION

Economic defense strategies derived from data, models and analysis about botnet attacker's economic behaviors, while respecting the target's own goals, represent a valuable addition to technical security measures. The use of game theory has the advantage of making explicit the trade-offs the target must make in its normal business models to accommodate and limit the effects of attacker's own business models. This extends statistical techniques [14] by further more allowing to design statistical strategies in a game theoretical framework. While this paper is inspired by a concrete case, the model can be extended to other types of service businesses, and shows explicitly how dynamics in the user-led demand can help the target.